\documentclass[twocolumn,showpacs,pre,floatfix,superscriptaddress]{revtex4-1}

\usepackage{color} 
\usepackage{tabularx} 
\usepackage{epsfig}
\usepackage{amsmath} 
\usepackage{amssymb} 
\usepackage{graphicx}
\usepackage{wasysym}
\usepackage{times}

\begin{document}

\title{Optimizing glassy $p$-spin models}

\author{Creighton K.~Thomas}
\affiliation {Department of Physics and Astronomy, Texas A\&M University,
College Station, Texas 77843-4242, USA}

\author{Helmut G.~Katzgraber} 
\affiliation {Department of Physics and Astronomy, Texas A\&M University,
College Station, Texas 77843-4242, USA}
\affiliation {Theoretische Physik, ETH Zurich, CH-8093 Zurich, Switzerland}

\date{\today}

\begin{abstract}

Computing the ground state of Ising spin-glass models with $p$-spin
interactions is, in general, an NP-hard problem.  In this work we show
that unlike in the case of the standard Ising spin glass with two-spin
interactions, computing ground states with $p=3$ is an NP-hard problem
even in two space dimensions.  Furthermore, we present generic exact
and heuristic algorithms for finding ground states of $p$-spin models
with high confidence for systems of up to several thousand spins.

\end{abstract}


\maketitle

\section{Introduction}
\label{sec:introduction}

Disordered materials, such as spin glasses, exhibit a rich equilibrium
and nonequilibrium behavior.  While the Edwards-Anderson Ising
spin-glass model \cite{edwards:75} incorporates the disorder and
frustration required to replicate glassy behavior \cite{young:98},
more generic models of disordered and glassy materials can provide
insight into a number of related problems. In particular, spin-glass
models with $p$-spin interactions have found a variety of applications
across disciplines. They are excellent examples of disordered model
systems in which the symmetry of the global states can be different
from that of the local degrees of freedom.

For example, the mean-field theory of $p$-spin models with $p
> 2$ is closely related to the behavior of structural glasses
\cite{kirkpatrick:87,kirkpatrick:87b,kirkpatrick:87c}.  The dynamics
of mean-field $p$-spin models has a close similarity to mode-coupling
theory \cite{goetze:92} for the dynamics of supercooled liquids: Both
the dynamical transition below which ergodicity breaking occurs and
the thermodynamic transition below which replica symmetry breaking
occurs (at the one-step level) \cite{bouchaud:98} can be found in
$p$-spin models \cite{bouchaud:98}.  Because the study of models of
interacting particles poses hard analytical and numerical challenges,
there have been many efforts in modeling structural glasses and
supercooled liquids using $p$-spin models \cite{larson:10}.

Similarly, there is a close relationship between implementations of
topologically-protected quantum computing and $p$-spin models with
disorder. To compute the error tolerance of topologically-protected
quantum computing proposals the problem is mapped onto a
statistical Ising spin-glass model with $p$-spin interactions
\cite{dennis:02}. The point in the disorder--temperature phase
diagram where the ferromagnetic--paramagnetic phase boundary
crosses the Nishimori line \cite{nishimori:81} represents the error
threshold---an important figure of merit---of the quantum computing
proposal.  For example, in the presence of bit-flip errors the Kitaev
proposal \cite{kitaev:03} with four-spin interactions maps onto a
two-dimensional random-bond Ising model (with $p=2$), the density of negative
bonds representing the density of bit-flip errors. Topological color
codes \cite{bombin:07b} instead map onto Ising spin-glass-like
Hamiltonians with three-spin interactions in the presence of bit-flip
errors \cite{bombin:08,katzgraber:09c}.

A common approach to better understand the low-temperature behavior of
spin glasses is to study in detail the structure of the ground state:
Zero-temperature optimization over the energetics of the system reveals
properties of the finite-temperature thermodynamics of the system.
This approach is complicated by the fact that these systems are,
in general, NP-hard. This means that they belong to a large class
of problems that are believed, in the worst case,
to be solvable only by investing
time exponential in the size of the problem \cite{garey:79} (e.g.,
the number of spins).  Elaborate techniques exist for solving NP-hard
spin-glass optimization problems \cite{liers:04}; however, there are special
cases, such as the
two-dimensional Ising spin glass with $p = 2$, that are not NP-hard,
and where exact efficient optimization is possible
\cite{barahona:82}.
Without disorder, the two-dimensional Ising model can be even
solved exactly \cite{onsager:44}, and techniques related to exact
solutions of the pure model have been directly useful for producing
efficient algorithms for simulating the two-dimensional Ising spin
glass as well \cite{saul:93,loh:06,thomas:09}.  Ground-state studies
of two-dimensional spin glasses with $p = 2$ have proven useful in
many aspects of spin-glass theory, including chaos \cite{bray:87},
reentrance \cite{amoruso:04}, and nonequilibrium behavior
\cite{thomas:08}.  Although the two-dimensional pure Ising model with
$p=3$ also permits an exact solution \cite{baxter:73,baxter:82},
no efficient simulation techniques are known for the corresponding
disordered problem.

Here we study the optimization of spin glasses with $p$-spin interactions. The
optimization problem of finding ground states of a generic spin-glass with
$p$-spin interactions is NP-hard.  In contrast to the two-dimensional Ising
spin glass with $p = 2$, we show here that even the special case of the
two-dimensional spin glass on any tripartite lattice with three-spin
interactions
is an NP-hard problem, at least for some disorder distributions. This is true
despite the existence of an exact solution for the pure case
\cite{baxter:73,baxter:82}.  The proof is based on a mapping of the {\em
three-dimensional} $p = 2$ Ising spin glass---which is known to be
NP-hard---onto the two-dimensional $p = 3$ Ising spin glass.  Nevertheless, we
present an approach that is capable of computing exact ground states of the
three-spin model for moderate-sized systems. While the exact approach presented
has been developed specifically for three-spin interactions, it can be
generalized to other values of $p$.  We also present a heuristic approach that
works quite well for systems of up to several thousand spins.  This technique
is general: The same code may be used to optimize a spin-glass problem with
{\em any geometry} and {\em any value} of $p$.  It consists of a genetic
algorithm using triadic crossover \cite{pal:94} combined with a local search.

In Sec.~\ref{sec:model_intro} we outline details of Ising models with
three-spin interactions, followed by a proof that the disordered three-spin Ising
model is NP-hard. We then present optimization techniques to study
models with $p$-spin interactions in Sec.~\ref{sec:opt} followed by
some results on test instances in Sec.~\ref{sec:tests}.

\section{Disordered three-spin Ising model}
\label{sec:model_intro}

The standard Edwards-Anderson (EA) spin-glass model with two-body
interactions is given by the Hamiltonian
\begin{eqnarray}
\mathcal{H}_\mathrm{EA} & = & -\sum_{\langle i j \rangle} J_{ij} s_i s_j,
\end{eqnarray}
where the sum is over all nearest-neighbor pairs $\langle i j \rangle$.
The Ising spins $s_i \in \{\pm 1\}$ interact via random couplings
$J_{ij}$.  The three-spin model, on the other hand, has spins placed
on the vertices of a triangulated lattice with plaquette interactions
$J_{ijk}$ between the spins  $i$, $j$ and $k$ on each plaquette
$\triangle_{ijk}$.  The Hamiltonian is
\begin{eqnarray}
\mathcal{H}_3 & = & -\sum_{\triangle_{ijk}} J_{ijk} s_i s_j s_k.
\label{eq:H3}
\end{eqnarray}
A plaquette is said to be unsatisfied when its contribution to the
Hamiltonian is positive. Typically, this model is studied either on
a triangular or a Union Jack lattice.

\begin{figure}
\includegraphics[width=\columnwidth]{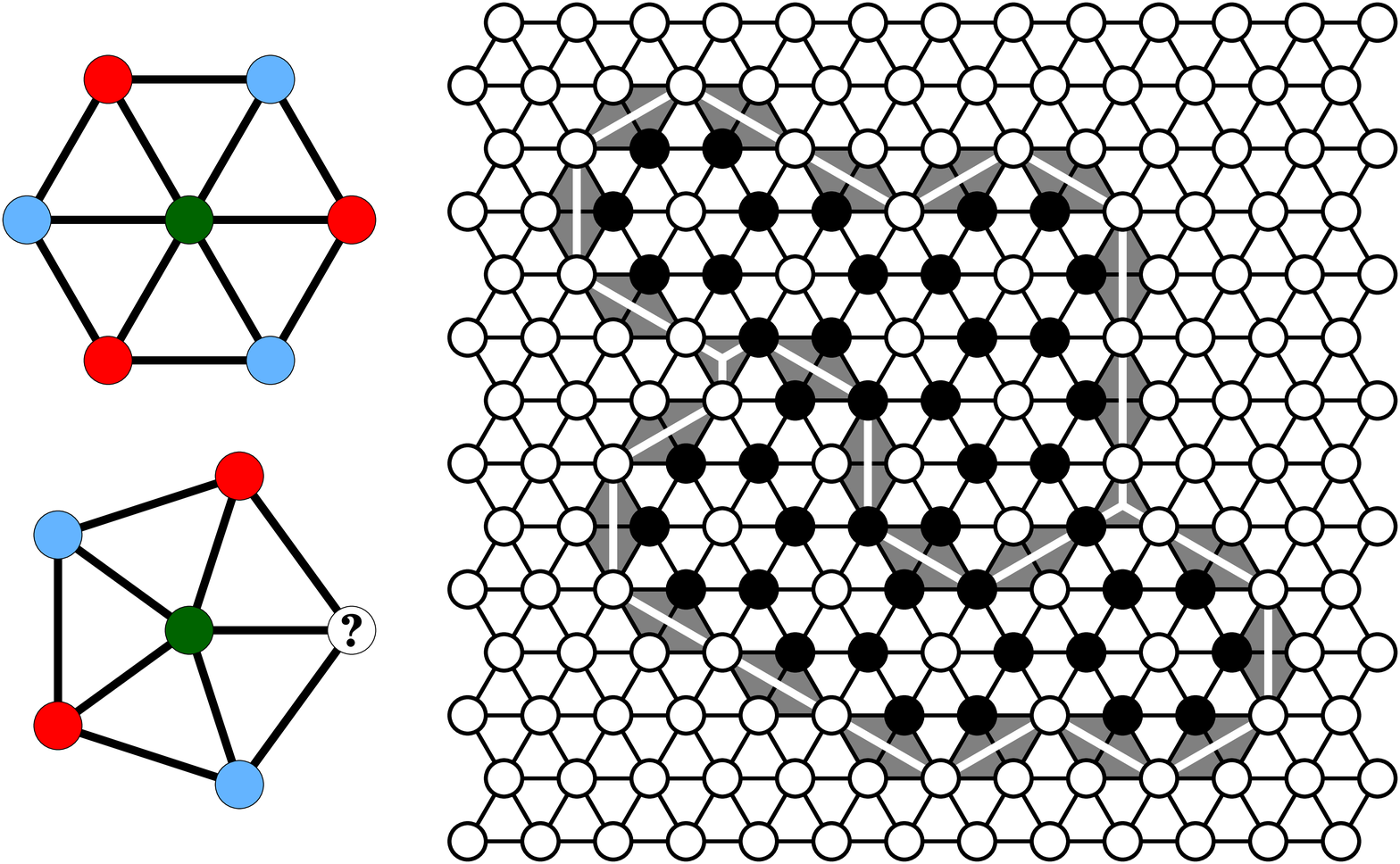}
\caption{(Color online) 
On a triangulated tripartite lattice each spin touches an even
number of triangular plaquettes (top left).  If a spin touches
an odd number of triangular plaquettes (bottom left), there is no
three-coloring of the graph, in contradiction with the assertion that
the graph be triTartite. Right panel: A section of a three-spin
system showing a domain wall separating states of the system.
White (black) circles show spins that are aligned (anti-aligned)
with some reference configuration. White plaquettes correspond to
terms that contribute identically in this new spin configuration as
in the reference configuration, while the gray plaquettes correspond
to terms with opposite sign to that of the reference configuration.
To guide the eye, the domain walls are highlighted with white lines,
separating two states of the system. Unlike in the Ising spin glass with $p = 2$,
three different states may come together at a point; thus the system
cannot be described entirely in terms of domain-wall loops.
}
\label{fig:tripartite}
\end{figure}

Both the triangular and Union Jack lattices are tripartite: One
can assign one of three colors to each site of the lattice such
that neighboring sites never have the same color; i.e., there are
three colored sublattices.  This model is most convenient to work
with on a tripartite lattice.  In this case, all spins border an even
number of plaquettes (see the left-hand side of Fig.~\ref{fig:tripartite}).
Furthermore, each pair of spins shares an even number of plaquettes (zero
or two), so the spins appear together in an even number of terms in the
Hamiltonian.
Flipping a spin therefore alters the
satisfaction of an even number of plaquettes adjacent to each spin.
All configurations of the system can be composed of individual spin flips, 
so the set of plaquettes
with differing satisfactions between any two spin configurations must
contain an even number of plaquettes touching each spin.  For a three-spin Ising
model on any tripartite lattice, this gives a
conservation rule: The parity of the number of unsatisfied plaquettes
touching each spin depends only on the instance of disorder.  While for
the EA Ising spin glass with $p = 2$ frustration properties are associated with the
plaquettes, in the three-spin model this conservation rule imparts
frustration properties to the {\em sites of the spin lattice}.
If the number of unsatisfied plaquettes touching some spin is even
for some configuration, then it is even for all spin configurations,
and if it is odd for some configuration, then it is frustrated in
that there is no configuration which has zero (or any even number of)
broken plaquettes touching this spin. In particular, if all spins
are unfrustrated, then the partition function is identical to that
of the pure system.

Each term of the Hamiltonian in Eq.~(\ref{eq:H3}) involves three
spins which are adjacent to one another, so each spin is a member of a
different color sublattice.  Unlike in standard spin-glass problems,
global spin-flip symmetry is absent: Flipping all spins negates the
Hamiltonian, rather than leaving it unchanged.  There is instead
a four-fold symmetry in the model from flipping all the spins
on certain colored sublattices.  Flipping all the spins on any
\emph{two} of the three colored sublattices leaves every term of the
Hamiltonian unchanged, so that the model is four-fold degenerate.
At low temperatures, these four degenerate states make up domains of
the system, with domain-wall excitations separating the pure state
regions. In the two-dimensional Ising spin glass with $p = 2$ \cite{thomas:07},
the boundaries between different states can be expressed entirely
in terms of the domain-wall loops.  In the three-spin model, a
similar loop description may be used to describe the domain-wall
separation between any two domains: The loops connect only sites on the
\emph{same} sublattice, and any plaquette the loop crosses is a member
of the domain wall.  The loop description is incomplete in this model,
because it is possible to have three different states come together at
a point (see the right-hand side of Fig.~\ref{fig:tripartite}). In this
sense, and because of the four-fold symmetry of the model, the Ising
model with three-spin interactions closely resembles a four-state
Potts model.  Despite the presence of domain walls that cannot be
classified as loops, a loop description of the problem will be a
useful limiting case to consider.  This loop description is helpful
for proving the three-spin ground-state problem is NP-hard and for
developing an optimization algorithm for this ground-state problem.

\section{The three-spin model is NP-hard}
\label{sec:np}

While instances of the bond-disordered Ising model with two-body interactions
in two dimensions may be solved exactly (i.e., find a ground-state
configuration or compute the partition function) with efficient
algorithms, the same model in three space dimensions is NP-hard
\cite{barahona:82}. We show here that the two-dimensional three-spin
Ising model is NP-hard as well by constructing a polynomial mapping by
which \emph{three-dimensional} spin-glass ground states may be found
using specially-constructed instances of the \emph{two-dimensional}
three-spin model.

First, we examine the Ising spin-glass model with $p = 2$. In two dimensions,
there is a one-to-one correspondence between spin configurations and
polygonal structures on the dual lattice, the sets of domain-wall
loops.  These polygons cross the bonds that are broken, relative to
some reference configuration \cite{thomas:07,thomas:09}. Therefore
each state of the Ising model corresponds to a state of the loop
model with the same energy, and the converse is also true: The models
are equivalent and have the exact same physical properties. In two
dimensions, the lowest-energy loop configurations may be found by
mapping the problem to a minimum-weight perfect matching problem on a
related (nonbipartite) lattice. Minimum-weight perfect matchings may be
solved efficiently using Edmonds's blossom algorithm \cite{edmonds:65}
(with subsequent fast implementations \cite{cook:99,kolmogorov:09}).

\begin{figure}

\includegraphics[width=\columnwidth]{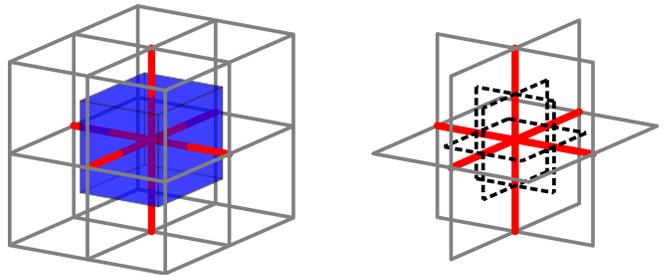}

\caption{(Color online) 
Simplest possible relative domain wall for the three-dimensional Ising spin
glass. On the left, one spin, in the middle, is flipped relative to a reference
configuration, changing the bonds which are drawn thicker (and shown in red),
and imposing the domain-wall polygon (the shaded cube in the center, here
blue).  The dashed lines shown on the right are the intersections of this
surface with the plaquettes (faces) of the cubic lattice.  This projects the
domain wall onto a wire-frame representation, which will be useful for mapping
onto the three-spin problem.  Only the plaquettes that intersect the domain
wall are shown. In general, each plaquette touches an even number of these
dashed edges, while each edge touches either zero or four of them.
}
\label{fig:cubetest}
\end{figure}

The correspondence between domain-wall configurations and spin
configurations is exact in {\em any} space dimension; i.e., a
similar construction may be made in general.  In three dimensions, the
domain-wall structures are polyhedra: sets of two-dimensional surfaces.
The energy of the system is given by the sum over all faces in each
set of polyhedra. Each face of the domain-wall polyhedra
crosses one edge from the spin lattice.
The faces of these polyhedra are defined on a polyhedral graph, another cubic lattice 
with one node at the center of each cube of the
spin graph.  The polyhedral graph is closely connected with the spin graph:
Each face of the polyhedral graph crosses one edge of the spin graph, while
each edge in the polyhedral graph is in turn crossed by one face of the spin
graph.
A wire-frame graph may also be defined that will be convenient for specifying
the set of domain walls that correspond to one spin configuration.  Each node
of this
wire-frame graph corresponds to either a face or an edge of the polyhedral
graph (all faces and edges are represented).  The graph is bipartite: The nodes
corresponding to a particular face (edge) of the domain wall graph are
connected to the nodes corresponding to the edges (faces) touching this face
(edge).
Note that, because the faces (edges) of the polyhedral graph correspond to the
edges (faces), this wire-frame graph is also defined on the faces and
edges of the spin graph.
A small subset of the polyhedral graph and the corresponding elements
of the wire-frame graph is shown, for example, in Fig.~\ref{fig:cubetest}.

We now specify the energetics of the spin system in terms of the domain walls
on the polyhedral graph.  Start by defining a reference spin
configuration ${r_i}$.  For each bond among nearest-neighbor pairs
$\langle i j \rangle$, let $R_{ij} \equiv r_i r_j$.  For a face $f$,
which crosses the bond between sites $i$ and $j$, let the weight $w$
be defined by $w(f) \equiv 2 J_{ij} R_{ij}$.  Then the Hamiltonian
may be rewritten
\begin{eqnarray}
\nonumber \mathcal{H} & = & -\sum_{\langle i j \rangle} 
J_{ij} \left( s_i s_j - R_{ij} + R_{ij} \right)\\
& = & \mathcal{H}_R + \mathcal{H}_P,
\end{eqnarray}
where $\mathcal{H}_R \equiv - \sum_{\langle i j \rangle} J_{ij}
R_{ij}$ is the constant contribution of the reference configuration
and $\mathcal{H}_P \equiv -\sum_{\langle i j \rangle} J_{ij} \left(
s_i s_j - R_{ij} \right) = \sum_{f \in P} w(f)$ is the contribution
from a polyhedral structure $P$ corresponding to the domain wall
separating configuration ${s_i}$ from ${r_i}$.  When $\mathcal{H}_P$
is minimized, so is $\mathcal{H}$.

\begin{figure}

\includegraphics[width=\columnwidth]{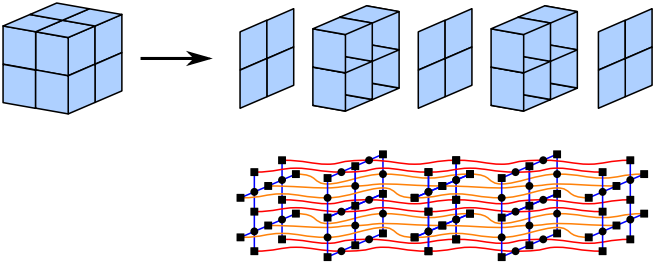}

\caption{(Color online) 
The three-dimensional Ising model on a cubic lattice (spins sit
at the vertices of the cubic lattice) is split into slices. The
intersection of a domain-wall surface with the faces and edges
of the graph uniquely defines the spin configuration. Below, the
graph is converted to a wire-frame form where the set of all spin
configurations is equated with the set of all selections of edges where
squares (corresponding to edges of the cubic lattice above) are type 1
junctions, which always touch either zero or four of the selected edges.
Circles---corresponding to faces of the cubic lattice above---are
type 2 junctions, which always touch an even number of selected edges.
The edges are allowed to cross (crossing edges correspond to a type
0 junction).  Colors are used to guide the eye.
}
\label{fig:cube}
\end{figure}

\begin{figure}

\includegraphics[width=\columnwidth]{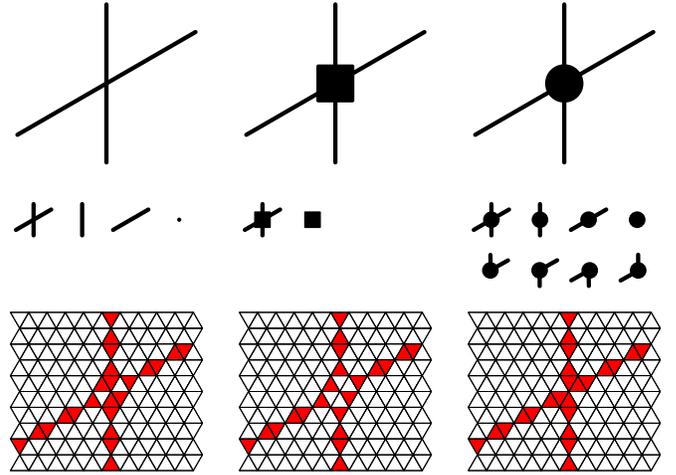}

\caption{(Color online) 
Junctions used in the mapping from the graph in Fig.~\ref{fig:cube}
to the three-spin model. Top: The junction type is shown as a crossing
with no symbol (type 0), square (type 1), and circle (type 2).  Below
the junctions is an enumeration of all possible edge selections for
each junction. Bottom: The junction is represented as a plaquette
disorder distribution. White plaquettes are given prohibitively
large weights such that they may not flip.  The only remaining spin
configurations correspond to flipping the plaquettes on paths according
to the rules above. The shaded (red) plaquettes all have zero weight, except
for one arbitrarily chosen plaquette in each square junction, which
is given the weight of the face of the polygon it corresponds to.
}
\label{fig:junctions} 
\end{figure}

These polyhedral structures may be uniquely defined by a wire-frame
representation: Here each face of the polyhedral graph is replaced by 
the intersections of the polyhedra with the faces of the
original spin lattice, the graph given by the spins and their interactions,
as is shown in Fig.~\ref{fig:cubetest}.
In this representation, each edge $e$ sits on a face $f_e$ of the
polyhedral structure $P$, and has weight $w(e) = w(f_e)/4$, so that
$\mathcal{H}_P = \sum_{f\in P} w(f) = \sum_{e\in P_e} w(e)$.  The set
of edges $P_e$ corresponding to the set of faces $P$ making up a
valid polyhedral set has two constraints.  First, when four edges
meet at the center of a single face on the polyhedral graph (at an
edge of the spin graph), then the face must as a whole be selected
or not, so either none of the edges is included, or all four are. We
call these ``type 1'' constraints. Second, when four edges meet at the
center of a face of the \emph{spin} graph, these are the edges on the
polyhedral graph, and any even number of these edges may be included
to give a valid polygon. We call these ``type 2'' constraints. In
Figs.~\ref{fig:cube}~and~\ref{fig:junctions}, the square junctions
follow the first constraint, while the circle junctions follow the
second constraint. This defines the set of all polyhedral structures
that are equivalent to the three-dimensional Ising spin glass model.

This wire-frame description may be drawn sliced into segments, as shown
in Fig.~\ref{fig:cube}; edges are allowed to cross, as is necessary
if one is to embed a three-dimensional graph in two dimensions.
Crossing edges must not interact with one another, introducing
one more junction (type 0).  Zero or four of the edges that come together may
be
included, or two edges opposite one another, but no turns are allowed.
This is shown in Fig.~\ref{fig:junctions}.

With three types of junctions, this wire-frame graph may be embedded in
the two-dimensional glassy three-spin model.  We take the wire-frame
graph, as drawn in Fig.~\ref{fig:cube}, and replicate each item of
the graph by setting the three-spin interactions $J_{ijk}$ of the
spin problem as appropriate. One of the colored sublattices 
(as defined in Sec. \ref{sec:model_intro}) of the three-spin
model is chosen to house all the domain walls in this graph.
This is enforced by setting bulk plaquette weights (i.e., the weights
of all plaquettes that are not allowed to be in a domain wall or in a
junction) to be prohibitively large, such that they must be satisfied
in the ground state.
In Fig.~\ref{fig:junctions}, these bulk plaquettes are colored white;
they make up the domains for which the domain-wall loops separate.
Each edge of the wire-frame graph is mapped to a zero-energy
domain-wall segment: a set of plaquettes with zero weight, but
surrounded by bulk plaquettes so that any valid spin configuration
has either all or none of the plaquettes unsatisfied (in the case of
a zero-weight plaquette, we choose to use the terms satisfied and
unsatisfied as though it had positive weight). Finally, the three
types of junctions are replaced by the three types of plaquette
``cities'' shown in Fig.~\ref{fig:junctions}, which are consistent
only with spin configurations that produce satisfactions in the
domain walls so that they interact with one another in exactly the
ways the edges in the wire-frame model are constrained to interact.
The weights of the plaquettes in the cities are all zero, except
in each square junction, where one plaquette is chosen to have the
weight of the face in the corresponding polyhedral model.  As either
all or none of the plaquettes in this junction are satisfied, it does
not matter which plaquette is given the nonzero weight.

The sum of the weights of all the bulk plaquettes does not depend
on the weights in the wire-frame model, so it is a constant,
$\mathcal{H}_c$.  The sum of the weights of the remaining (domain
wall and junction) plaquettes depends only on the satisfaction of
each square junction, as all other plaquettes have zero weight.
Therefore the Hamiltonian $\mathcal{H}_3$ of the three-spin model
with these plaquette weights may be written in terms of the polygonal
structure $P$ as
\begin{eqnarray}
\mathcal{H}_3 & = & \mathcal{H}_c + \sum_{f \in P} w(f) - 
			\sum_{f \notin P} w(f) .
\end{eqnarray}
Calling $\mathcal{H}_f = \sum_f w(f)$ a disorder-dependent constant
related to $\mathcal{H}_R$, we obtain 
\begin{eqnarray}
\mathcal{H}_3 & = \mathcal{H}_c + \mathcal{H}_f + 2 \mathcal{H}_P
\end{eqnarray}
such that 
\begin{eqnarray}
\mathcal{H} & = \mathcal{H}_R - \frac{1}{2}\mathcal{H}_c
		- \frac{1}{2}\mathcal{H}_f + \frac{1}{2} \mathcal{H}_3 .
\end{eqnarray}
Because $\mathcal{H}_R$, $\mathcal{H}_c$, and $\mathcal{H}_f$ are
all constants that may be computed efficiently for each instance of
the disorder, this directly relates the ground state of a specific
instance of the two-body three-dimensional Ising spin glass to a specific
instance of the two-dimensional three-spin model.

The number of spins necessary in the three-spin model is polynomial in
the number of spins in the three-dimensional Ising spin-glass problem.
This is because the only constraint forcing the number of spins in
the three-spin model to scale faster than linearly with the number
of spins in the three-dimensional Ising spin glass model is that a
new junction must be created for each crossed edge. If we assume the
worst scaling possible, i.e., that the length of every domain wall
must scale linearly with the total number of domain walls in the
problem, then the number of spins necessary in the three-spin model
is bounded from above by $\mathcal{O}(N^2) \times \mathcal{O}(N^2)
= \mathcal{O}(N^4)$, for $N$ spins in the three-dimensional Ising
spin glass, which is still polynomial. As the ground state of any
three-dimensional two-body Ising spin glass may be found by solving an instance
of the two-dimensional three-spin model ground-state problem with a
number of spins polynomial in $N$, finding the ground state of this
three-spin model is NP-hard.

\section{Optimization methods for $p$-spin models}
\label{sec:opt}

The three-spin spin-glass model is NP-hard, so it is unsurprising that
we have not found any fast (i.e., polynomial run time with system size)
exact algorithm for computing ground states.  We start by describing
an exact integer linear program (ILP) technique for optimizing the
three-spin glassy model using cutting planes. This technique scales
exponentially with the system size to find exact results, but it is
particularly useful because one may quickly find tight lower bounds
on the ground-state energy. We then present an efficient heuristic
technique for arbitrary $p$ using the triadic crossover genetic update
of Pal \cite{pal:94}, with which we can solve, with high confidence,
systems with discrete disorder up to several thousand spins.

\subsection{Cutting plane technique (exact algorithm)}

Ground states of the Ising spin glass have been computed using
ILP approaches \cite{barahona:88,liers:04}.  Here we show that the
three-spin model may also be optimized exactly by mapping the problem
onto an ILP combined with a cutting-plane technique.  This approach does not
depend on the distribution of disorder, and its performance is roughly
comparable for Gaussian and bimodal disorder.

An ILP may be expressed in canonical form with coefficient vectors $c$
and $b$, coefficient matrix $A$, and vector $x$ of integer variables as
\begin{eqnarray}
\textrm{Minimize } && c^T x \\
\textrm{Subject to } && A x \leq b . \nonumber
\end{eqnarray}
The function to minimize, $c^T x$, is called the objective function;
the elements of $A x \leq b$ are the constraint inequalities. The
problem is specified by giving the values of all elements of $A$, $b$,
and $c$. The function $c^T x$ is, up to a linear transformation, equal
to the Hamiltonian to be minimized, while each row of the constraint
inequality equation is a constraint on the values of the plaquettes
designed to enforce that only valid spin configurations are allowed
(the number of rows of $A$ and the number of elements of $b$ is the
number of constraints in the linear program). An optimal vector $x$
gives the lowest value of the objective function among all possible
$x$ that satisfy the constraints. This problem is often posed as
a maximization problem; this is equivalent, as one may replace $c$
with $-c$ and solve the maximization problem. Typical linear program
solvers can optimize in either direction.

Solving an ILP is an NP-hard problem, but linear programs without the
integer constraints permit efficient solutions, e.g., by simplex or
interior point methods \cite{chvatal:83}.  It is therefore possible
to solve ILPs by successively adding cutting planes to linear programs:
Additional constraints are
added to enforce that the solution vector $x$ contains only integers.
This can be achieved by constructing a tight convex hull around the
permitted values with intersections only at permitted integer values.
This tight hull requires exponentially many constraints to specify,
so in practice one tries to find and add only the most important
extra constraints.

For both the two-body Ising spin glass and the three-spin spin-glass model, the
Hamiltonian is not linear in the spin degrees of freedom. Therefore,
expressing the problem as an ILP requires additional work.  One may
perform a change of variables such that the Hamiltonian is linear in
the new variables.  For the Ising spin-glass problem, the Hamiltonian
is quadratic in the spin degrees of freedom, so the Hamiltonian
is a weighted sum of edge satisfactions $e_{ij} \equiv s_i s_j$.
The spin values are uniquely defined by the edge satisfactions, up to
a global spin-flip degeneracy. In the three-spin case, the Hamiltonian
is linear in three-spin plaquette terms $x(\ell) \equiv s_i s_j s_k$
where plaquette $\ell$ touches spins $i$, $j$ and $k$. The vectors
$x$ and $c$ therefore are of size $N_p$, the number of plaquettes in
the system, and for each plaquette $\ell$ defined as above, $c(\ell)
= J_{ijk}$.  Performing a linear transformation on the plaquette
satisfaction variables from $x(\ell) \in \left\{-1,1\right\}\rightarrow
x(\ell) \in \left\{0,1\right\}$ optimizes the new Hamiltonian
$\mathcal{H}_{\mathrm{LP}} = (\mathcal{H} + N_p)/2$, for which
the optimization problem is equivalent, and which trivially maps
back to the original problem. The spin values may be extracted by
changing variables back to the original spin degrees of freedom;
the configuration is determined only up to the four-fold degeneracy
of the model, so two spins (preferably adjacent to one another)
may be assigned randomly, and this forces the values of all other spins.

As in the spin-glass case, the change of variables to reduce
the cubic program to a linear program moves the complexity of
the optimization problem from the cubic objective function, which
produces a very complex energy landscape (but with no constraints
besides the variables being integers), to a linear objective function,
which has additional constraints to ensure that the two formulations
are equivalent. The extra constraints are necessary because not all
plaquette configurations correspond to a valid spin configuration.
The constraints to be added are a generalization of the odd-cycle
(OC) constraints used in the spin-glass technique.

The parity conservation rule introduced in Sec.~\ref{sec:model_intro}
implies a related principle: The number of plaquettes whose
satisfaction differ between any two spin configurations must be even.
This can be seen by changing from one spin configuration to the
other one spin flip at a time.  No spin flip can change the parity,
so it is always even. Furthermore, the interaction of any loop on one of the
colored lattices
(using the representation in Fig.~\ref{fig:tripartite}) with any spin
configuration
has the same parity constraint.  In the bulk of the sample, this
constraint is implied by the previous one, but when the system has
periodic boundaries, it adds the case of system-spanning loops, which
do not correspond to spin configurations. The inclusion of these
system-spanning loops ensures that the solution is in the correct
topological sector, which is necessary for the spin configuration to
be well defined when converting from plaquette values to spin values
(c.f.~the extended ground state in spin glasses \cite{thomas:07}).

These parity constraints lead to OC-like inequalities.  Let
$\mathcal{C}$ be a set of objects for which the vector $x$ has $N_p$ elements
given by the union of the set of all spin-configuration
differences with the set of all loop differences (as defined in
Sec.~\ref{sec:model_intro})
with a spin
configuration. These objects can be defined by the vector $x$ which gives
(possibly fractional) plaquette satisfactions.
For each member $C \in \mathcal{C}$, all $F
\subset C$ with $\left| F \right|$ odd satisfy
\begin{eqnarray}
\label{eq:OC}
x(F) - x(C \setminus F) \leq \left| F \right| - 1.
\end{eqnarray}
Each such equation rules out the case where $\forall \ell \in F,
x(\ell) = 1 $ and $0$ otherwise -- with $|F|$ odd, this is not allowed
by the parity constraint.  Thus this set of constraint inequalities
provides the cutting planes to eliminate invalid solutions. Clearly,
the number of constraints is huge: Every possible spin configuration
contributes many constraints of this type to the linear program, so
it is unreasonable to include them all.  It is therefore necessary
to find the most important constraints without which the solution is
incorrect and ignore as many others as possible, i.e., such that $A$
and $b$ are not too large. If too few constraints are included at a
given step, the solution given by the linear program at that step will
not correspond to a valid spin configuration. It will have energy lower
than the ground-state energy because the problem is underconstrained
(adding new constraints can only raise the value of the solution).

Given a test solution where the linear program solution contains
either odd-plaquette violations or non-integer variables (typically
both), one searches for new constraints that are violated by the
current configuration. These are added to the problem, and the
linear program is solved again. This is repeated until the result
is a valid plaquette configuration, in which case all constraints
are satisfied (including the integer variables condition).  Also,
for discrete disorder distributions, the solution is complete if the
lower bound given by this technique is close enough to confirm that
a heuristic solution is an exact solution.

If adding constraints does not produce a solution to the ILP, one
may also branch: Assign the values of one or more variables, and
search given these variables. The full solution requires exhausting
the exponentially-many possibilities, but some of these possibilities
can be eliminated if both good upper and lower bounds can be computed
for the cases. Many practical frameworks exist for combining branching
with cutting planes. For the implementation of the algorithms described
here, we use the Coin Branch and Cut (CBC) framework with the Coin
Linear Program Solver (CLP) simplex algorithm to solve individual
linear programs \cite{coin}.

We outline the procedures used for finding new constraints step-by-step:

\begin{itemize}

\item[$\Box$]{\textit{Local spin-flip constraint finder} --- The
simplest sets of constraints involve $C$ being the six plaquettes
adjacent to a given spin. With $2^{|C|-1} = 32$ possible choices of
$F$, all possible constraint violations may be tested around each
spin, although this is typically unnecessary: It is often simple
to identify which combinations are most likely and test only those
(for example, in the case where all weights are currently integers,
adding the $x=1$ cases and subtracting the $x=0$ cases is the only
one of the 32 which can produce a violated constraint).  The simplest
generalization of the local spin-flip constraint finder is taking the
plaquettes that change when flipping multiple nearby spins. We have
tried all combinations of up to four spins. These help the convergence
of the ILP only marginally. Therefore they are only included as a
last resort check if all other constraint finders fail.}

\item[$\Box$]{\textit{Loop constraint finder} --- In analogy
with the constraint finders for the Ising spin glass presented in
Ref.~\onlinecite{barahona:88}, one can use all the finders used
for loops in the Ising spin glass on the loops of the tripartite
{\em sublattices}.  The major difference is that each edge in the
loop description of the three-spin model contains two plaquettes.
This actually simplifies some aspects of the computation because all
loops are guaranteed to have even length. Two constraint finders,
the spanning tree heuristic for odd cycles (SHOC) and the shortest
paths exact finder, odd-cycle (OC) in Ref.~\onlinecite{barahona:88},
are particularly useful for our application, although any constraint
finder from the spin-glass problems could be similarly ported to the
three-spin problem. Some of these constraint finders will naturally
produce some even-cycle violated ``constraints'' that are not valid
because all constraints must contain an odd number of items in $F$.
These are normally discarded, but it is useful to store them for
later use in the genetic constraint finder below.}

\item[$\Box$]{\textit{``Worm'' constraint finder} --- All the
constraint finders for the ILP solution of the Ising spin glass work
with loops; this is not the only kind of constraints available in
the three-spin model; i.e., another class of finders is also needed.
One approach employed to find nonloop constraints is to do a search
by flipping adjacent spins successively.  At each step of the search,
we keep the set of flipped spins and the direction in which the set of
spins (the ``worm'') is growing.  Four possibilities are considered:
capping the worm and testing if it produces a valid constraint, or
letting the worm grow straight or turn to the left or right.  If the
worm is to grow, the weights of the two new plaquettes are recorded.
For each plaquette $\ell$, if $x(\ell) < 0.5$, one adds $x(\ell)$
to the total weight so far, otherwise add $1-x(\ell)$.  If the
total weight so far ever exceeds $1$, the search may stop because
no constraint found after this point can satisfy a constraint in the
form of Eq.~(\ref{eq:OC}).}

\item[$\Box$]{\textit{``Genetic'' constraint finder} --- If the above
constraint finders are insufficient, we have developed one additional
powerful constraint finder. Any spin configurations may give valid
constraints, and finding the tightest convex hull is very difficult,
even with the highly-effective constraint finders outlined above.
Additional constraints may be found by combining sets of plaquettes
from different constraint inequalities using a symmetric difference.
In the case where the two constraint inequalities contain at least
one common variable, the new inequalities are not simply a linear
combination of the two previous ones, so they exclude a different
region of parameter space and may be useful.  It is particularly
helpful when the cases where $|F|$ is even are kept in the above
constraint finders, because one is most likely to find a constraint
inequality with $|F|$ odd when combining an odd case with an even case.
In practice, these often produce new constraint inequalities that
are violated.  We call this a genetic constraint finder because it
takes the population of currently known constraint inequalities,
all of which might be satisfied by the current configuration, and it
generates new constraint inequalities (better offspring) by combining
two constraints at random. This constraint finder would also allow
one to find new constraints in ILP solutions of the Ising spin glass.}
\end{itemize}

The exact technique presented here depends on finding good constraint
inequalities; the machinery for finding these inequalities is more developed
for quadratic programming, so we have developed a partial reduction technique
to take advantage of this (the loop constraint finder) in addition to our
constraint finders, which work directly with the three-spin problem.  This
reduction is highly influenced by the geometry of the problem, which makes it
effective for finding good constraints.  Buchheim and Rinaldi have also
recently developed a different technique that fully reduces a cubic
programming problem to quadratic program \cite{buchheim:07}.  This reduction
does not take advantage of known geometrical constraints, but it has the
substantial advantage that it would eliminate the need for the more specialized
constraint finders, allowing one to use only quadratic programming techniques
to solve the problem.  It would be interesting to compare these two methods for
solving this ground-state problem.

\subsection{Local search optimization}

We describe a simple generic local search algorithm. It is similar to
standard local search optimization methods \cite{martin:04}.  While it
is not particularly effective for continuous disorder distributions
(in that case a variable-depth search performs better), it works
quite well for the case of discrete disorder.  This local search
algorithm has been designed with the three-spin problem in mind,
but it is generic: It works well for all the spin systems
we have tried when they have discrete energy levels, regardless of
space dimension or the value of $p$. It consists of a depth-first
search where at each step in the search a spin is test-flipped and
the search may overcome energy barriers up to some cutoff energy. It
is most easily implemented with a boolean recursive function. In all,
$N$ searches are run, one starting from each spin in the system, in
random order, for a given cutoff search depth $d_\mathrm{max}$ and
energy barrier $E_\mathrm{max}$ to overcome. One of these searches
is implemented by calling, for site $i$, \texttt{searchstep}$(0,i)$,
where this is the boolean function defined by \\

\noindent
boolean \textbf{function} \texttt{searchstep}$(d,i)$\\
\indent \textbf{if} $d > d_\mathrm{max}$\\ 
\indent\indent \textbf{return} false\\
\indent flip $s_i \gets -s_i$\\
\indent \textbf{if} $E(\{s_i\}) <= E_\mathrm{max}$\\
\indent\indent \textbf{if} $E(\{s_i\}) <= E_\mathrm{t}$\\
\indent\indent\indent \textbf{return} true\\
\indent\indent \textbf{for each} $j$ which neighbors $i$\\
\indent\indent\indent  \textbf{if} \texttt{searchstep}$(d+1,j)$ returns true\\
\indent\indent\indent\indent \textbf{return} true\\
\indent reset $s_i \gets -s_i$\\
\indent \textbf{return} false\\

\noindent When the function returns true, the energy has been lowered
by switching to a new spin configuration. When it returns false, no
change has been made to the spin configuration.  For bimodal disorder,
this procedure typically finds the true ground state in small systems
of up to $\sim 300$ spins ($L \le 18$ in two space dimensions),
when $d_\mathrm{max} > L$ and $E_\mathrm{max}$ is given by twice the smallest
energy increment in the system.  It is also a useful search technique
for the genetic algorithm described in the next Section.

\subsection{Genetic algorithm with local search}
\label{sec:genetic}

Genetic algorithms are useful heuristic techniques for solving
optimization problems with complex energy landscapes.  A genetic
algorithm consists of a population of solutions---many distinct
instances of the problem that eventually evolve toward the solution of
the optimization problem. One way for a genetic algorithm to proceed
is that at each step of the algorithm parent instances are chosen
and reproduced: The offspring (or child instances) are generated by
combining the parent solutions in some way and the children are added
to the population. Some members of the population are eliminated
according to a fitness criterion to keep the population size from
growing.

In order for a genetic algorithm to be effective, there must be
an efficient mechanism for reproduction: Child solutions must be
as fit as their parents, or they are likely to be eliminated soon,
although there must be enough variation such that child solutions
are not simply repeats of previous members of the population. One
effective reproduction mechanism for spin systems is triadic crossover
\cite{pal:95,pal:96}.  Like in standard crossover reproduction, the
bits of two children are created by swapping bits of two parents. In
this case, which parent's bit goes to which child is decided by
comparing one of the parents with a third parent. When the spin values
in parents 1 and 3 are equal, child 1 inherits the spin value from
parent 1, while child 2 inherits the spin value from parent 2.  When the
spin values differ between parents 1 and 3, child 1 inherits the spin
value from parent 2, and child 2 inherits the spin values from parent 1.

Other than the Hamiltonian, which spatially couples the spins, the
triadic crossover technique does not use the spatial structure (good
regions of spins to flip are chosen solely from their correlation
among different instances) so storing each spin as one bit and using
bitwise operations on strings of bits significantly decreases both
the storage space necessary and the number of operations necessary
to perform the steps of this algorithm.

Pal originally used the triadic crossover technique with only very
simple randomized local optimization techniques.  Triadic crossover
has also been exploited in conjunction with highly-sophisticated
optimization procedures to find heuristic ground states in the
three-dimensional Ising spin glass up to $14^3$ \cite{hartmann:97}.
We employ an intermediate approach: The local optimization algorithm
from the previous subsection is quite simple but performs very well.
At each step of the genetic algorithm we perform a triadic crossover
reproduction to produce two child instances from three randomly
chosen parents.  Each of these child instances is then optimized
by a single local search sweep (starting once from each spin in the
system, with a depth cutoff of $L$).  Then, each child is compared
against a randomly chosen instance in the population with fitness
below the median (energy above the median of the whole population).
If the child's fitness is better, it may replace this randomly chosen
instance. To keep the population heterogeneous, we allow this
replacement only if the child is not a repeat of any current member
of the population. No additional randomization is carried out in this
procedure.  This is adequate for some cases (such as the Ising spin glass
results presented below), while in other cases it helps to
carry out parallel evolution starting from several different initial
populations to increase heterogeneity.

We have produced a highly portable code which is effective for systems
of up to several thousand spins (performance is similar to that of
the highly-sophisticated code in Ref.~\onlinecite{hartmann:97}).
The simplicity of this technique makes it particularly convenient to
port to different types of interactions: We can use the \emph{same
code} to optimize the two-dimensional and three-dimensional Ising
spin glass, the Sherrington-Kirkpatrick (mean field) Ising spin glass,
as well as for arbitrary $p$-spin models.

\subsection{Combining the techniques}

While the exact solution of the three-spin problem using the ILP
solution is quite time consuming for more than $\sim 300$ spins,
the cutting planes technique quickly provides an exact lower bound
on the ground-state energy that is often very close to the true
ground state. It is common to use heuristic solutions as a part of
a branch-and-cut technique \cite{liers:04}; in cases of discrete
disorder, in particular, the ILP lower bound is quite commonly below
the heuristic solution by less than the energy of a single spin flip.
The heuristic ground state solution is therefore shown to be exact.
For a system with $3$-spin interactions and $24^2$ spins, we find
that the solution can be confirmed exact in a reasonable amount of
time in $95\%$ of the samples. In the other cases, it is likely that
the heuristic solution is still correct, but we have not proven it
with this technique.

\section{Genetic algorithm results}
\label{sec:tests}

This algorithm is intended for use on $p$-spin models for any $p$, but
it is difficult to test its performance in these models because there
are no exact techniques known for the optimization of large instances
of $p$-spin models. To study the performance of the algorithm, we
therefore use the two-dimensional Ising spin glass with $p = 2$, followed by some
tests on the three-spin case. We then compute the ground-state energy
per spin $E_0 / L^2$ for the disordered Ising model with $p=3$ on a
triangular lattice.

\subsection{Benchmark case: Ising spin glass ($p = 2$)}

\begin{figure}
\includegraphics[width=\columnwidth]{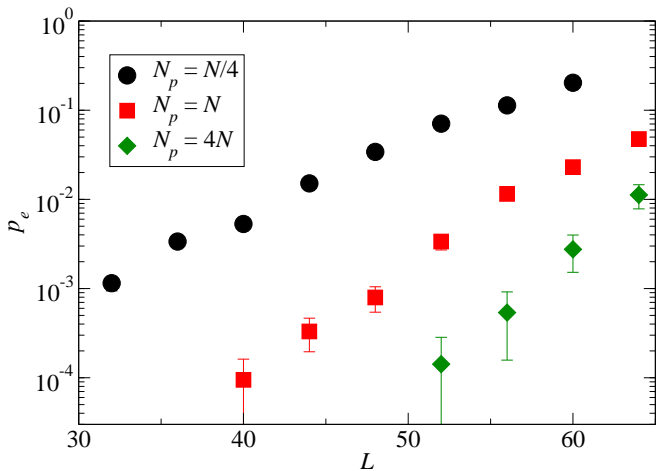}
\caption{(Color online) 
Probability $p_e$ that the genetic algorithm described in
Sec.~\ref{sec:genetic} gives an incorrect ground state for the
two-dimensional Ising spin glass with $\pm J$ disorder and $p = 2$.
Increasing the population size allows for more genetic variation,
such that the algorithm is able to find the true ground states of
large systems.  No parallel runs are performed to further increase
variation in this case.  Where the statistical error bars are not
visible, they are smaller than the symbols.
}
\label{fig:p_e}
\end{figure}

Fast algorithms for optimizing the two-dimensional Ising spin
glass with $p = 2$ are readily available \cite{barahona:82,thomas:07,pardella:08,liers:10}, so we can
find the parameters for which the genetic algorithm does produce
correct solutions with high confidence.  We use the Ising spin
glass with bimodal interactions; bond values are chosen to be $1$
or $-1$ with equal probability.  For simplicity, we compare against
cases where the ground state is equal to the extended ground state
\cite{thomas:07}; these ground states are expected to behave as
the other ground states, so that this choice should not affect the
performance results significantly.  For a given population size $N_p$,
we find the probability that the genetic algorithm gives an incorrect
result, as shown in Fig.~\ref{fig:p_e}.  For $N_p \geq N$, with $N$
being the number of spins, these errors are very rare up to rather
large system sizes; for $N_p = 4 N$, we failed to find any errors
for $L < 50$ in $2\times 10^4$ attempts, while for $N_p = N$, the
same is true for $L < 40$.

\begin{figure}
\includegraphics[width=\columnwidth]{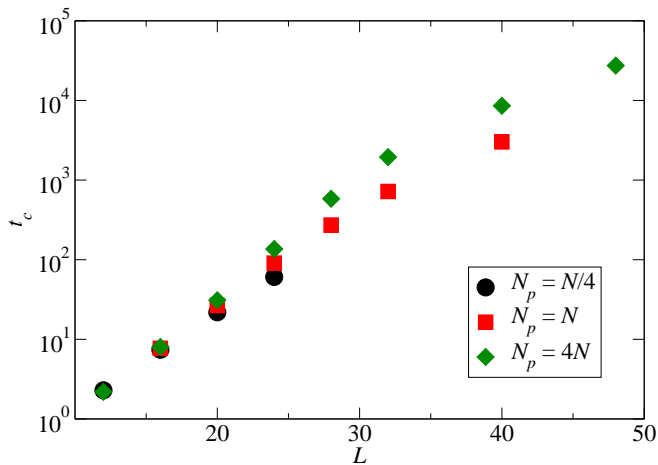}
\caption{(Color online) 
Number of steps $t_c$ necessary to find the lowest-energy state
(in the cases where the true ground state is actually found)
for the two-dimensional Ising spin glass with $\pm J$ disorder.
If more than one error has occurred at a given system size $L$ (see
Fig.~\ref{fig:p_e}), the solution technique is presumed ineffective
at $L$ and the data point is omitted.  The time required to find
the correct result scales exponentially with $L$ for the system
sizes studied, however with a small enough prefactor to make these
computations reasonable up to $L\approx50$.  Statistical error bars
are smaller than the symbols.
}
\label{fig:t_c}
\end{figure}

For the two-dimensional Ising spin glass, the local search algorithm is
quite effective, so for small systems, it takes few updates to find the
true ground state. For larger systems the number of updates necessary
to reach the true ground state in a two-dimensional Ising spin glass
with $p = 2$ scales exponentially (or at least as a power law with
exponent $>6$). In Fig.~\ref{fig:t_c}, the number of reproduction steps
necessary to reach the true ground state is plotted in the cases where
the true ground state can actually be reached; when the true ground
state is unreachable for some samples, we exclude the run-time data.

\subsection{The disordered Ising model with $p = 3$}

\begin{figure}
\includegraphics[width=\columnwidth]{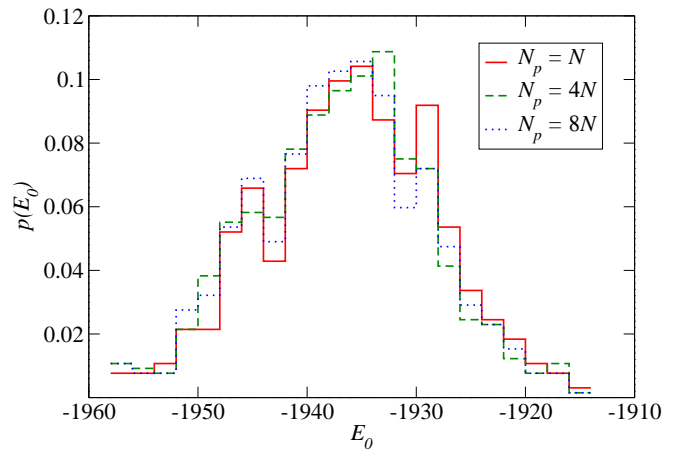}
\caption{(Color online) 
Ground-state energy $E_0$ histogram showing the fraction of samples at
each energy, for the disordered three-spin Ising model on a triangular
lattice with $L=36$. Variation among the population sizes shows that
the true ground state is not reached in every case, although the
centers of the distributions are very close together and shapes of
the distribution are similar.
}
\label{fig:3spinhist}
\end{figure}

Because the algorithm outlined in Sec.~\ref{sec:genetic} is intended
for cases where $p > 2$, we investigate the disordered Ising model
with $p=3$ on a two-dimensional triangular lattice.  The plaquette
energies $J_{ijk}$ are chosen independently to be $1$ or $-1$ with
equal probability.  With the integer linear program technique detailed
above, the genetic algorithm is seen to successfully find ground
states with high probability up to $L\approx24$, but beyond this system
size we have no exact check of the technique. Performance tests are
therefore much more difficult to perform. In Fig.~\ref{fig:3spinhist}
a histogram of the ground-state energy probability is given for $L=36$
for population sizes $N_p = N, 4N, 8N$.  The fluctuations in the
rate of occurrence of each ground-state energy $p(E_0)$ show that the
convergence is not exact even at this moderate system size.  However,
the distribution shifts only slightly as the population size changes,
implying that the distribution is converging to the exact result.

\begin{figure}
\includegraphics[width=\columnwidth]{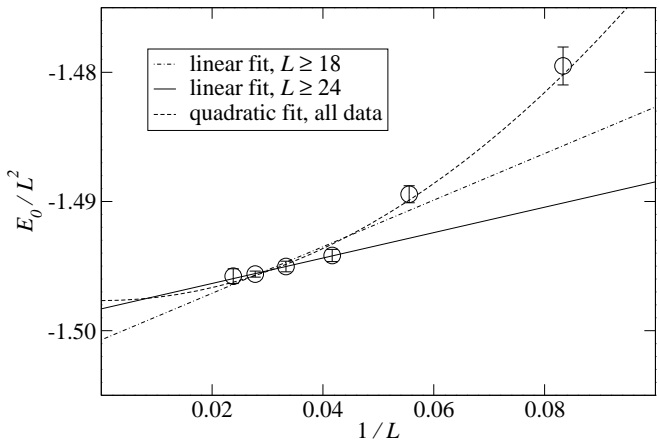}
\caption{
Ground-state energy density, $E_0/L^2$ of the disordered Ising
model with $p=3$ on a triangular lattice.  The plaquette interactions are
bimodally distributed with positive and negative interactions being equally
probable.  To extrapolate to the thermodynamic limit,
we performed linear and quadratic fits, as shown.  Our estimate of
the ground-state energy density in this model is $-1.499(3)$.
}
\label{fig:gs_e}
\end{figure}

Finally, we estimate the ground-state energy density of the
two-dimensional spin glass with $p = 3$ on a triangular lattice.
The ground-state energy $E_0$ [the lowest energy possible in Eq.~(\ref{eq:H3})]
is computed for 400 samples for each
of $L=12$, $18$, $24$, $30$, $36$, and $42$.  Based on the convergence
of the ground-state energy histogram, the error is dominated by
the statistical fluctuations among samples: Any systematic error
in the average is expected to be less than this statistical error.
To extrapolate to the thermodynamic limit we plot the energy density
at each system size $L$, $E_0/L^2$ as a function of $1/L$ and take
the limit as $1/L\rightarrow 0$, as shown in Fig.~\ref{fig:gs_e}.
Because the system sizes are moderate, finite-size effects can be seen in
the data. In order to fairly extrapolate to $L\rightarrow \infty$,
we perform linear and quadratic curve fits, varying which system
sizes to include in the curve fits. Two example linear curve fits are
shown in Fig.~\ref{fig:gs_e}: The solid line includes $L \ge 24$, whereas
the dot-dashed line includes $L \ge 18$.
The quadratic fit, which includes all $L$ values shown, gives a
similar value for the thermodynamic extrapolation.  
From these fits,
we estimate that the ground-state energy density in the disordered
Ising model with $p=3$ on a triangular lattice is $-1.499(3)$.

\section{Conclusions}

We have shown that the optimization of a disordered Ising model
with three-spin interactions is an NP-hard problem in two dimensions,
so that efficient exact algorithms are not expected to exist, even
for simple cases of glassy $p$-spin models.  Optimization of NP-hard
problems is a difficult task: We map this problem onto an integer
linear program problem and can find exact ground states using a
branch-and-cut technique.  This works well in small systems, but takes
time exponential in the size of the system.  To better address physical
problems in $p$-spin models, heuristic approaches are important.
We present an effective heuristic algorithm that combines a genetic
approach with local search optimization to give ground states with high
confidence in systems of up to several thousand spins.  This algorithm
is simple and general: Our implementation can work for any geometry and
with any spin interactions.  These techniques should prove useful for
future work on the low-temperature glassy dynamics of $p$-spin models.

\begin{acknowledgments} 

H.G.K.~acknowledges support from the Swiss National Science Foundation
(Grant No.~PP002-114713). The authors acknowledge Texas A\&M University
for access to their hydra cluster, the Texas Advanced Computing Center
(TACC) at the University of Texas at Austin for providing HPC resources
(Ranger Sun Constellation Linux Cluster) and ETH Zurich for CPU time
on the Brutus cluster.

\end{acknowledgments}

\vspace{-1em}
\bibliography{refs,comments}

\begin{thebibliography}{41}
\expandafter\ifx\csname natexlab\endcsname\relax\def\natexlab#1{#1}\fi
\expandafter\ifx\csname bibnamefont\endcsname\relax
  \def\bibnamefont#1{#1}\fi
\expandafter\ifx\csname bibfnamefont\endcsname\relax
  \def\bibfnamefont#1{#1}\fi
\expandafter\ifx\csname citenamefont\endcsname\relax
  \def\citenamefont#1{#1}\fi
\expandafter\ifx\csname url\endcsname\relax
  \def\url#1{\texttt{#1}}\fi
\expandafter\ifx\csname urlprefix\endcsname\relax\def\urlprefix{URL }\fi
\providecommand{\bibinfo}[2]{#2}
\providecommand{\eprint}[2][]{\url{#2}}

\bibitem[{\citenamefont{Edwards and Anderson}(1975)}]{edwards:75}
\bibinfo{author}{\bibfnamefont{S.~F.} \bibnamefont{Edwards}} \bibnamefont{and}
  \bibinfo{author}{\bibfnamefont{P.~W.} \bibnamefont{Anderson}},
  \bibinfo{journal}{J. Phys. F: Met. Phys.} \textbf{\bibinfo{volume}{5}},
  \bibinfo{pages}{965} (\bibinfo{year}{1975}).

\bibitem[{\citenamefont{Young}(1998)}]{young:98}
\bibinfo{editor}{\bibfnamefont{A.~P.} \bibnamefont{Young}}, ed.,
  \emph{\bibinfo{title}{Spin Glasses and Random Fields}}
  (\bibinfo{publisher}{World Scientific}, \bibinfo{address}{Singapore},
  \bibinfo{year}{1998}).

\bibitem[{\citenamefont{Kirkpatrick and
  Wolynes}(1987{\natexlab{a}})}]{kirkpatrick:87}
\bibinfo{author}{\bibfnamefont{T.~R.} \bibnamefont{Kirkpatrick}}
  \bibnamefont{and} \bibinfo{author}{\bibfnamefont{P.~G.}
  \bibnamefont{Wolynes}}, \bibinfo{journal}{Phys. Rev. B}
  \textbf{\bibinfo{volume}{36}}, \bibinfo{pages}{8552}
  (\bibinfo{year}{1987}{\natexlab{a}}).

\bibitem[{\citenamefont{Kirkpatrick and
  Wolynes}(1987{\natexlab{b}})}]{kirkpatrick:87b}
\bibinfo{author}{\bibfnamefont{T.~R.} \bibnamefont{Kirkpatrick}}
  \bibnamefont{and} \bibinfo{author}{\bibfnamefont{P.~G.}
  \bibnamefont{Wolynes}}, \bibinfo{journal}{Phys. Rev. A}
  \textbf{\bibinfo{volume}{35}}, \bibinfo{pages}{3072}
  (\bibinfo{year}{1987}{\natexlab{b}}).

\bibitem[{\citenamefont{Kirkpatrick and Thirumalai}(1987)}]{kirkpatrick:87c}
\bibinfo{author}{\bibfnamefont{T.~R.} \bibnamefont{Kirkpatrick}}
  \bibnamefont{and}
  \bibinfo{author}{\bibfnamefont{D.}~\bibnamefont{Thirumalai}},
  \bibinfo{journal}{Phys. Rev. B} \textbf{\bibinfo{volume}{36}},
  \bibinfo{pages}{5388} (\bibinfo{year}{1987}).

\bibitem[{\citenamefont{G{\"o}tze and Sj{\"o}gren}(1992)}]{goetze:92}
\bibinfo{author}{\bibfnamefont{W.}~\bibnamefont{G{\"o}tze}} \bibnamefont{and}
  \bibinfo{author}{\bibfnamefont{L.}~\bibnamefont{Sj{\"o}gren}},
  \bibinfo{journal}{Rep. Prog. Phys.} \textbf{\bibinfo{volume}{55}},
  \bibinfo{pages}{241} (\bibinfo{year}{1992}).

\bibitem[{\citenamefont{Bouchaud et~al.}(1998)\citenamefont{Bouchaud,
  Cugliandolo, Kurchan, and M\'ezard}}]{bouchaud:98}
\bibinfo{author}{\bibfnamefont{J.-P.} \bibnamefont{Bouchaud}},
  \bibinfo{author}{\bibfnamefont{L.~F.} \bibnamefont{Cugliandolo}},
  \bibinfo{author}{\bibfnamefont{J.}~\bibnamefont{Kurchan}}, \bibnamefont{and}
  \bibinfo{author}{\bibfnamefont{M.}~\bibnamefont{M\'ezard}}, in
  \emph{\bibinfo{booktitle}{Spin glasses and random fields}}, edited by
  \bibinfo{editor}{\bibfnamefont{A.~P.} \bibnamefont{Young}}
  (\bibinfo{publisher}{World Scientific}, \bibinfo{address}{Singapore},
  \bibinfo{year}{1998}).

\bibitem[{\citenamefont{Larson et~al.}(2010)\citenamefont{Larson, Katzgraber,
  Moore, and Young}}]{larson:10}
\bibinfo{author}{\bibfnamefont{D.}~\bibnamefont{Larson}},
  \bibinfo{author}{\bibfnamefont{H.~G.} \bibnamefont{Katzgraber}},
  \bibinfo{author}{\bibfnamefont{M.~A.} \bibnamefont{Moore}}, \bibnamefont{and}
  \bibinfo{author}{\bibfnamefont{A.}~\bibnamefont{Young}},
  \bibinfo{journal}{Phys. Rev. B} \textbf{\bibinfo{volume}{81}},
  \bibinfo{pages}{064415} (\bibinfo{year}{2010}).

\bibitem[{\citenamefont{Dennis et~al.}(2002)\citenamefont{Dennis, Kitaev,
  Landahl, and Preskill}}]{dennis:02}
\bibinfo{author}{\bibfnamefont{E.}~\bibnamefont{Dennis}},
  \bibinfo{author}{\bibfnamefont{A.}~\bibnamefont{Kitaev}},
  \bibinfo{author}{\bibfnamefont{A.}~\bibnamefont{Landahl}}, \bibnamefont{and}
  \bibinfo{author}{\bibfnamefont{J.}~\bibnamefont{Preskill}},
  \bibinfo{journal}{J. Math. Phys.} \textbf{\bibinfo{volume}{43}},
  \bibinfo{pages}{4452} (\bibinfo{year}{2002}).

\bibitem[{\citenamefont{{Nishimori}}(1981)}]{nishimori:81}
\bibinfo{author}{\bibfnamefont{H.}~\bibnamefont{{Nishimori}}},
  \bibinfo{journal}{Prog. Theor. Phys.} \textbf{\bibinfo{volume}{66}},
  \bibinfo{pages}{1169} (\bibinfo{year}{1981}).

\bibitem[{\citenamefont{Kitaev}(2003)}]{kitaev:03}
\bibinfo{author}{\bibfnamefont{A.~Y.} \bibnamefont{Kitaev}},
  \bibinfo{journal}{Ann. Phys.} \textbf{\bibinfo{volume}{303}},
  \bibinfo{pages}{2} (\bibinfo{year}{2003}).

\bibitem[{\citenamefont{Bombin and Martin-Delgado}(2007)}]{bombin:07b}
\bibinfo{author}{\bibfnamefont{H.}~\bibnamefont{Bombin}} \bibnamefont{and}
  \bibinfo{author}{\bibfnamefont{M.~A.} \bibnamefont{Martin-Delgado}},
  \bibinfo{journal}{Phys. Rev. Lett.} \textbf{\bibinfo{volume}{98}},
  \bibinfo{pages}{160502} (\bibinfo{year}{2007}).

\bibitem[{\citenamefont{Bombin and Martin-Delgado}(2008)}]{bombin:08}
\bibinfo{author}{\bibfnamefont{H.}~\bibnamefont{Bombin}} \bibnamefont{and}
  \bibinfo{author}{\bibfnamefont{M.~A.} \bibnamefont{Martin-Delgado}},
  \bibinfo{journal}{Phys. Rev. A} \textbf{\bibinfo{volume}{77}},
  \bibinfo{pages}{042322} (\bibinfo{year}{2008}).

\bibitem[{\citenamefont{Katzgraber et~al.}(2009)\citenamefont{Katzgraber,
  Bombin, and Martin-Delgado}}]{katzgraber:09c}
\bibinfo{author}{\bibfnamefont{H.~G.} \bibnamefont{Katzgraber}},
  \bibinfo{author}{\bibfnamefont{H.}~\bibnamefont{Bombin}}, \bibnamefont{and}
  \bibinfo{author}{\bibfnamefont{M.~A.} \bibnamefont{Martin-Delgado}},
  \bibinfo{journal}{Phys. Rev. Lett.} \textbf{\bibinfo{volume}{103}},
  \bibinfo{pages}{090501} (\bibinfo{year}{2009}).

\bibitem[{\citenamefont{Garey and Johnson}(1979)}]{garey:79}
\bibinfo{author}{\bibfnamefont{M.~R.} \bibnamefont{Garey}} \bibnamefont{and}
  \bibinfo{author}{\bibfnamefont{D.~S.} \bibnamefont{Johnson}},
  \emph{\bibinfo{title}{Computers and intractability: a guide to the theory of
  NP-completeness}} (\bibinfo{publisher}{Freeman}, \bibinfo{address}{San
  Francisco}, \bibinfo{year}{1979}).

\bibitem[{\citenamefont{Liers et~al.}(2004)\citenamefont{Liers, J{\"u}nger,
  Reinelt, and Rinaldi}}]{liers:04}
\bibinfo{author}{\bibfnamefont{F.}~\bibnamefont{Liers}},
  \bibinfo{author}{\bibfnamefont{M.}~\bibnamefont{J{\"u}nger}},
  \bibinfo{author}{\bibfnamefont{G.}~\bibnamefont{Reinelt}}, \bibnamefont{and}
  \bibinfo{author}{\bibfnamefont{G.}~\bibnamefont{Rinaldi}}, in
  \emph{\bibinfo{booktitle}{New Optimization Algorithms in Physics}}, edited by
  \bibinfo{editor}{\bibfnamefont{A.~K.} \bibnamefont{Hartmann}}
  \bibnamefont{and} \bibinfo{editor}{\bibfnamefont{H.}~\bibnamefont{Rieger}}
  (\bibinfo{publisher}{Wiley-VCH}, \bibinfo{address}{Berlin},
  \bibinfo{year}{2004}).

\bibitem[{\citenamefont{Barahona}(1982)}]{barahona:82}
\bibinfo{author}{\bibfnamefont{F.}~\bibnamefont{Barahona}},
  \bibinfo{journal}{J. Phys. A} \textbf{\bibinfo{volume}{15}},
  \bibinfo{pages}{3241} (\bibinfo{year}{1982}).

\bibitem[{\citenamefont{Onsager}(1944)}]{onsager:44}
\bibinfo{author}{\bibfnamefont{L.}~\bibnamefont{Onsager}},
  \bibinfo{journal}{Phys. Rev.} \textbf{\bibinfo{volume}{65}},
  \bibinfo{pages}{117} (\bibinfo{year}{1944}).

\bibitem[{\citenamefont{Saul and Kardar}(1993)}]{saul:93}
\bibinfo{author}{\bibfnamefont{L.}~\bibnamefont{Saul}} \bibnamefont{and}
  \bibinfo{author}{\bibfnamefont{M.}~\bibnamefont{Kardar}},
  \bibinfo{journal}{Phys. Rev. E} \textbf{\bibinfo{volume}{48}},
  \bibinfo{pages}{R3221} (\bibinfo{year}{1993}).

\bibitem[{\citenamefont{Loh and Carlson}(2006)}]{loh:06}
\bibinfo{author}{\bibfnamefont{Y.~L.} \bibnamefont{Loh}} \bibnamefont{and}
  \bibinfo{author}{\bibfnamefont{E.~W.} \bibnamefont{Carlson}},
  \bibinfo{journal}{Phys. Rev. Lett.} \textbf{\bibinfo{volume}{97}},
  \bibinfo{pages}{227205} (\bibinfo{year}{2006}).

\bibitem[{\citenamefont{Thomas and Middleton}(2009)}]{thomas:09}
\bibinfo{author}{\bibfnamefont{C.~K.} \bibnamefont{Thomas}} \bibnamefont{and}
  \bibinfo{author}{\bibfnamefont{A.~A.} \bibnamefont{Middleton}},
  \bibinfo{journal}{Phys. Rev. E} \textbf{\bibinfo{volume}{80}},
  \bibinfo{pages}{046708} (\bibinfo{year}{2009}).

\bibitem[{\citenamefont{Bray and Moore}(1987)}]{bray:87}
\bibinfo{author}{\bibfnamefont{A.~J.} \bibnamefont{Bray}} \bibnamefont{and}
  \bibinfo{author}{\bibfnamefont{M.~A.} \bibnamefont{Moore}},
  \bibinfo{journal}{Phys. Rev. Lett.} \textbf{\bibinfo{volume}{58}},
  \bibinfo{pages}{57} (\bibinfo{year}{1987}).

\bibitem[{\citenamefont{Amoruso and Hartmann}(2004)}]{amoruso:04}
\bibinfo{author}{\bibfnamefont{C.}~\bibnamefont{Amoruso}} \bibnamefont{and}
  \bibinfo{author}{\bibfnamefont{A.~K.} \bibnamefont{Hartmann}},
  \bibinfo{journal}{Phys. Rev. B} \textbf{\bibinfo{volume}{70}},
  \bibinfo{pages}{134425} (\bibinfo{year}{2004}).

\bibitem[{\citenamefont{Thomas et~al.}(2008)\citenamefont{Thomas, White, and
  Middleton}}]{thomas:08}
\bibinfo{author}{\bibfnamefont{C.~K.} \bibnamefont{Thomas}},
  \bibinfo{author}{\bibfnamefont{O.~L.} \bibnamefont{White}}, \bibnamefont{and}
  \bibinfo{author}{\bibfnamefont{A.~A.} \bibnamefont{Middleton}},
  \bibinfo{journal}{Phys. Rev. B} \textbf{\bibinfo{volume}{77}},
  \bibinfo{pages}{092415} (\bibinfo{year}{2008}).

\bibitem[{\citenamefont{Baxter and Wu}(1973)}]{baxter:73}
\bibinfo{author}{\bibfnamefont{R.~J.} \bibnamefont{Baxter}} \bibnamefont{and}
  \bibinfo{author}{\bibfnamefont{F.~Y.} \bibnamefont{Wu}},
  \bibinfo{journal}{Phys. Rev. Lett.} \textbf{\bibinfo{volume}{31}},
  \bibinfo{pages}{1294} (\bibinfo{year}{1973}).

\bibitem[{\citenamefont{Baxter}(1982)}]{baxter:82}
\bibinfo{author}{\bibfnamefont{R.}~\bibnamefont{Baxter}},
  \emph{\bibinfo{title}{{Exactly Solved Models in Statistical Mechanics}}}
  (\bibinfo{publisher}{Academic Press}, \bibinfo{address}{London},
  \bibinfo{year}{1982}).

\bibitem[{\citenamefont{Pal}(1994)}]{pal:94}
\bibinfo{author}{\bibfnamefont{K.~F.} \bibnamefont{Pal}}, in
  \emph{\bibinfo{booktitle}{Parallel Problem Solving from Nature}}
  (\bibinfo{publisher}{Springer}, \bibinfo{address}{Berlin},
  \bibinfo{year}{1994}), p. \bibinfo{pages}{170}.

\bibitem[{\citenamefont{Thomas and Middleton}(2007)}]{thomas:07}
\bibinfo{author}{\bibfnamefont{C.~K.} \bibnamefont{Thomas}} \bibnamefont{and}
  \bibinfo{author}{\bibfnamefont{A.~A.} \bibnamefont{Middleton}},
  \bibinfo{journal}{Phys. Rev. B} \textbf{\bibinfo{volume}{76}},
  \bibinfo{pages}{220406(R)} (\bibinfo{year}{2007}).

\bibitem[{\citenamefont{Edmonds}(1965)}]{edmonds:65}
\bibinfo{author}{\bibfnamefont{J.}~\bibnamefont{Edmonds}},
  \bibinfo{journal}{Canad. J. Math.} \textbf{\bibinfo{volume}{17}},
  \bibinfo{pages}{449} (\bibinfo{year}{1965}).

\bibitem[{\citenamefont{Cook and Rohe}(1999)}]{cook:99}
\bibinfo{author}{\bibfnamefont{W.}~\bibnamefont{Cook}} \bibnamefont{and}
  \bibinfo{author}{\bibfnamefont{A.}~\bibnamefont{Rohe}},
  \bibinfo{journal}{INFORMS J. Comp.} \textbf{\bibinfo{volume}{11}},
  \bibinfo{pages}{138} (\bibinfo{year}{1999}).

\bibitem[{\citenamefont{Kolmogorov}(2009)}]{kolmogorov:09}
\bibinfo{author}{\bibfnamefont{V.}~\bibnamefont{Kolmogorov}},
  \bibinfo{journal}{Math. Prog. Comp.} \textbf{\bibinfo{volume}{1}},
  \bibinfo{pages}{43} (\bibinfo{year}{2009}).

\bibitem[{\citenamefont{Barahona et~al.}(1988)\citenamefont{Barahona,
  Gr{\"o}tschel, J{\"u}nger, and Reinelt}}]{barahona:88}
\bibinfo{author}{\bibfnamefont{F.}~\bibnamefont{Barahona}},
  \bibinfo{author}{\bibfnamefont{M.}~\bibnamefont{Gr{\"o}tschel}},
  \bibinfo{author}{\bibfnamefont{M.}~\bibnamefont{J{\"u}nger}},
  \bibnamefont{and} \bibinfo{author}{\bibfnamefont{G.}~\bibnamefont{Reinelt}},
  \bibinfo{journal}{Oper. Res.} \textbf{\bibinfo{volume}{36}},
  \bibinfo{pages}{493} (\bibinfo{year}{1988}).

\bibitem[{\citenamefont{Chv\'{a}tal}(1983)}]{chvatal:83}
\bibinfo{author}{\bibfnamefont{V.}~\bibnamefont{Chv\'{a}tal}},
  \emph{\bibinfo{title}{Linear Programming}} (\bibinfo{publisher}{Freeman},
  \bibinfo{address}{New York}, \bibinfo{year}{1983}).

\bibitem[{coi()}]{coin}
\bibinfo{note}{Coin-or, \url{http://www.coin-or.org/}}.

\bibitem[{\citenamefont{Buchheim and Rinaldi}(2007)}]{buchheim:07}
\bibinfo{author}{\bibfnamefont{C.}~\bibnamefont{Buchheim}} \bibnamefont{and}
  \bibinfo{author}{\bibfnamefont{G.}~\bibnamefont{Rinaldi}},
  \bibinfo{journal}{SIAM J. Optim.} \textbf{\bibinfo{volume}{18}},
  \bibinfo{pages}{1398} (\bibinfo{year}{2007}).

\bibitem[{\citenamefont{Martin}(2004)}]{martin:04}
\bibinfo{author}{\bibfnamefont{O.~C.} \bibnamefont{Martin}}, in
  \emph{\bibinfo{booktitle}{New Optimization Algorithms in Physics}}, edited by
  \bibinfo{editor}{\bibfnamefont{A.~K.} \bibnamefont{Hartmann}}
  \bibnamefont{and} \bibinfo{editor}{\bibfnamefont{H.}~\bibnamefont{Rieger}}
  (\bibinfo{publisher}{Wiley-VCH}, \bibinfo{address}{Berlin},
  \bibinfo{year}{2004}).

\bibitem[{\citenamefont{Pal}(1995)}]{pal:95}
\bibinfo{author}{\bibfnamefont{K.~F.} \bibnamefont{Pal}},
  \bibinfo{journal}{Biol. Cybern.} \textbf{\bibinfo{volume}{73}},
  \bibinfo{pages}{335} (\bibinfo{year}{1995}).

\bibitem[{\citenamefont{Pal}(1996)}]{pal:96}
\bibinfo{author}{\bibfnamefont{K.~F.} \bibnamefont{Pal}},
  \bibinfo{journal}{Physica A} \textbf{\bibinfo{volume}{223}},
  \bibinfo{pages}{283} (\bibinfo{year}{1996}).

\bibitem[{\citenamefont{Hartmann}(1997)}]{hartmann:97}
\bibinfo{author}{\bibfnamefont{A.}~\bibnamefont{Hartmann}},
  \bibinfo{journal}{Europhys. Lett.} \textbf{\bibinfo{volume}{40}},
  \bibinfo{pages}{429} (\bibinfo{year}{1997}).

\bibitem[{\citenamefont{Pardella and Liers}(2008)}]{pardella:08}
\bibinfo{author}{\bibfnamefont{G.}~\bibnamefont{Pardella}} \bibnamefont{and}
  \bibinfo{author}{\bibfnamefont{F.}~\bibnamefont{Liers}},
  \bibinfo{journal}{Phys. Rev. E} \textbf{\bibinfo{volume}{78}},
  \bibinfo{pages}{056705} (\bibinfo{year}{2008}).

\bibitem[{\citenamefont{Liers and Pardella}(2010)}]{liers:10}
\bibinfo{author}{\bibfnamefont{F.}~\bibnamefont{Liers}} \bibnamefont{and}
  \bibinfo{author}{\bibfnamefont{G.}~\bibnamefont{Pardella}},
  \bibinfo{journal}{Computational Optimization and Applications}
  p.~\bibinfo{pages}{1} (\bibinfo{year}{2010}).

\end{thebibliography}

\end{document}